# Electro-Mechanical Response of Top-Gated LaAlO$_3$/SrTiO$_3$ Heterostructures

Feng Bi,[1] Mengchen Huang,[1] Chung Wung Bark,[2,*] Sangwoo Ryu,[2] Sanghan Lee,[2] Chang-Beom Eom,[2] Patrick Irvin[1] and Jeremy Levy[1,†]

[1]*Dept. of Physics & Astronomy, University of Pittsburgh, Pittsburgh, Pennsylvania, 15260, USA*

[2]*Dept. of Materials Science, University of Wisconsin-Madison, Madison, Wisconsin, 53706, USA*

[*] Present address: Gachon University, Korea

[†] To whom correspondence should be addressed: jlevy@pitt.edu

LaAlO$_3$/SrTiO$_3$ heterostructures are known to exhibit a sharp, hysteretic metal-insulator transition (MIT) with large enhanced capacitance near depletion. To understand the physical origin of this behavior, the electromechanical response of top-gated LaAlO$_3$/SrTiO$_3$ heterostructures is probed using two simultaneous measurement techniques: piezoforce microscopy (PFM) and capacitance spectroscopy. PFM measurements reveal local variations in the hysteretic response, which is directly correlated with capacitance measurements. The enhanced capacitance at the MIT is linked to charging/discharging dynamics of nanoscale conducting islands, which are revealed through PFM imaging and time-resolved capacitance and piezoresponse measurements.

# I. INTRODUCTION

Oxide heterostructures have received increased attention due to the variety of exhibited behavior[1-7] (insulating, metallic, ferroelectric, ferromagnetic and superconducting) as well as their potential for nanoelectronic device applications[8-13]. The interface between LaAlO$_3$ (LAO) and TiO$_2$-terminated SrTiO$_3$ (STO) is *n*-type conducting[1] beyond a critical LAO thickness of 3 unit cells[3] and can be tuned through the metal-insulator transition (MIT) using a metallic top gate[4,14-16], back gate[3] or electrolyte gate[17]. Li *et al.*[13] found that near the MIT, the capacitance of gated LAO/STO heterostructures is enhanced significantly (over 40%) beyond the geometric value: $C_{geom} = \frac{\varepsilon_{LAO} \cdot A}{d}$, where $\varepsilon_{LAO}$ is the dielectric permittivity of LAO, and *A* and *d* are the cross section and the spacing of the capacitor, respectively. Bark *et al.* have characterized gated LAO/STO/LSAT structures using piezoelectric force microscopy (PFM) and observed hysteretic behavior near the MIT[18]. However, the intrinsic origin of the top-gate-tuned MIT and the mechanism for capacitance enhancement as well as PFM hysteresis are still not well understood.

To help clarify the physical origin of these transport and mechanical effects, *simultaneous* capacitance-voltage (CV) spectroscopy and PFM measurements are performed on top-gated LAO/STO structures. The CV measurements measure the ability of the interface to store electric charge under various top-gate bias conditions. The abrupt change in capacitance, and associated in-plane resistance change, is associated with the MIT. PFM is a powerful tool that senses the local surface distortion of the LAO/STO heterostructure with high precision, yielding an electromechanical response either on specified locations or over the desired areas. The piezoresponse signal is believed to arise from a carrier-mediated Jahn-Teller effect near the LAO/STO interface. Simultaneously measured CV and local piezoresponse measurements yield new insights into local charging/discharging dynamics at mesoscopic scales near the MIT. The time scale for charging/discharging dynamics can be obtained by time-resolved piezoresponse measurements; the observed time scales agree well with previously reported frequency dispersion of the capacitance enhancement phenomena.

## II.   SAMPLE AND DEVICE FABRICATION

LAO/STO heterostructures are fabricated by depositing 5 u.c. and 12 u.c. LAO films on $TiO_2$-terminated (001) STO substrates using pulsed laser deposition with *in situ* high pressure reflection high energy electron diffraction (RHEED). Before deposition, low-miscut (<0.1°) STO substrates are etched using buffered HF acid to keep the $TiO_2$-termination. Then the STO substrates are annealed at 1000 °C for 2~12 hours so the atomically flat surfaces are created. During the deposition, the KrF exciter laser ($\lambda = 248$ nm) beam is focused on a stoichiometric LAO single crystal target with energy density 1.5 mJ/cm$^2$ and each LAO unit cell is deposited by 50 laser pulses. Two different growth conditions are used: (1) the substrate growth temperate T=550°C and chamber background partial oxygen pressure $P=10^{-3}$ mbar; (2) the substrate growth temperate $T$=780°C and chamber background $P=10^{-5}$ mbar. For samples grown in condition (2), after deposition, they are annealed at 600 °C in 300 mbar oxygen pressure environment for one hour to reduce oxygen vacancies.

The capacitor devices are designed to have a geometry similar to that reported in Ref. 13. The top-gate electrodes are circular in shape, with a diameter ranging between 100 μm and 600 μm. The electrodes connecting the interface are arch-shaped with a width 20 μm and fixed separation of 50 μm to the edge of the circular top gates. These arch-shaped electrodes surrounding the top gate are intended to provide a more isotropic electric field under applied bias.

The capacitor electrodes are deposited on the LAO/STO samples via DC- sputtering. As shown in **FIG. 1**(a), the arch-shaped electrodes are prepared by first creating 25-nm trenches via Ar-ion milling, followed by deposition of 4 nm Ti and 30 nm Au. Then a series of metallic circular top gates (4 nm Ti and 40 nm Au) are deposited on the LAO surface. The whole sample is affixed to a ceramic chip carrier using silver paint. Electrical contacts to the device are made with an ultrasonic wire bonder, using gold wires. Current-voltage (*I-V*) measurements are performed between arch shaped electrodes on all devices; measurements indicate (e.g., **FIG. 1**(b)) that the contacts have no Schottky barriers, similar to other reports[3, 7, 14, 17].

Experimental measurements from three devices whose parameters are summarized in TABLE I. For comparison between measurements and with results from the published literature, most of the results shown are from Device A. Similar results have been obtained for each of the three devices investigated. Selected additional results on other devices are shown in the Appendix.

TABLE I.  Device Parameters.

|  | Top-gate diameter | LAO thickness | Growth condition |
|---|---|---|---|
| Device A | 100 µm | 12 u.c. | Grow at 780°C, partial oxygen pressure $P=10^{-5}$ mbar, annealed |
| Device B | 500 µm | 5 u.c. | Grow at 550°C, partial oxygen pressure $P=10^{-3}$ mbar |

| Device C | 600 μm | 5 u.c. | Grow at 780°C, partial oxygen pressure $P=10^{-5}$ mbar, annealed |

## III. EXPERIMENTAL RESULTS

### A. Top-gate tuned MIT at interface

To check whether the top gate can effectively tune the carriers at interface, a simple transport measurement of the interface is performed between two electrodes etched to the interface. The experiment setup is illustrated in **FIG. 2**(a). An ac source voltage with amplitude $V_{ac} = 0.1$ V and frequency $f = 37.3$ Hz is applied to one arch-shaped electrode. The ac drain current $I$ is measured by a lock-in amplifier through the second arch shaped electrode. The circular top gate is connected with a dc bias $V_g$ to tune the MIT at the interface. To avoid large current, a series resistance $R_r = 115$ M ohm is used in the circuit. Therefore the resistance of interface under $V_g$ can be obtained from $R = V_{ac}/I - R_r$, which is plotted in **FIG. 2**(b). The sharp hysteretic change in resistance of nearly two orders of magnitude coincides with the sharp changes in capacitance and indicates a MIT near $V_g=0$. Note that the resistance does not become immeasurably large in the insulating phase because the gate does not extend to the gap region between the two arch electrodes.

### B. Capacitance and piezoresponse measurements setup

Simultaneous PFM and CV measurements are performed using the experimental configuration illustrated in **FIG. 3**(a). A combined dc and "dual-frequency ac signal" is applied to the interface $V_{interface}(t) = -V_{dc} - V_{PFM}\cos(2\pi F_{PFM}t) - V_C\cos(2\pi F_C t)$; the top electrode is held at virtual ground in order to eliminate possible electrostatic interactions with the atomic force microscope (AFM) probe. Note that the dc component applied to the interface to be -$V_{dc}$, so the equivalent top gate bias is $V_{dc}$. Adjusting $V_{dc}$ tunes the LAO/STO interface between conducting and insulating phases. In order to measure the PFM response, an AFM tip is placed in contact with the top-gate electrode. The AFM tip deflection at the excitation frequency $F_{PFM}$ is detected using a lock-in amplifier.

The capacitance as a function of gate bias $C(V_{dc})$ is measured simultaneously using a capacitance bridge similar to that described in Ref. 13. The capacitance bridge circuit is shown as the green circuit in **FIG. 3**(a). The sample is excited with an ac voltage $V_C$ in one arm. In another arm, a synchronized ac voltage $V_b$ with the same frequency and 180-degree phase shift is applied to a standard capacitor $C_s$ (typically 10 pF). The signal at the balancing point is measured by a second lock-in amplifier. During the capacitance measurement, the phase and amplitude of $V_C$ are held constant, while $V_b$ is adjusted in a feedback loop to null the capacitance channel (out-of-phase or "Y" channel) at the balancing point. The sample capacitance is obtained from the equation: $C = C_s V_b / V_C$. Both capacitance and piezoresponse measurements are performed in an enclosed AFM chamber with a small amount of scattered red light ($\lambda \geq 620\ nm$), which does not have an observable effect on transport properties.

Fixed-frequency CV curves are also acquired using a commercial capacitance bridge (Andeen-Hagerling 2500A) with an ac excitation voltage given by $V_C$=10 mV, $F_c$=1 kHz and $V_{dc}$ sweep at a rate $dV_{dc}/dt$=0.067 V/s. Capacitance measurements (**FIG. 3**(b)) indicate a MIT as the gate bias $V_{dc}$ is varied. When $V_{dc}$ decreases ($V_{dc}$<2V), the capacitance signal begins to drop from its saturated value. As $V_{dc}$ decreases further ($V_{dc}$<0V), the capacitance reaches a low and steady base level, indicating that the electron density underneath the circular top gate is substantially depleted by the electric field. The depleted interface can be restored to a conducting state by increasing $V_{dc}$. As $V_{dc}$ increases and becomes positive ($V_{dc}>0$), the capacitance increases several-fold, indicating a conducting channel is restored at the interface. The capacitance variation during gate tuning is comparable to that reported in Refs. 13-14. The *CV* curve in **FIG. 3**(b) exhibits a stable hysteretic behavior, in agreement with previous studies[14, 16].

### C. Piezoresponse measurements

Local piezoresponse measurements are also performed on the same devices (**FIG. 3**(c)). The piezoresponse signal is measured at a fixed location on the top electrode as a function of $V_{dc}$. The ac excitation voltage is given by $V_{PFM}=20$ mV and $F_{PFM}=293$ kHz, with a voltage sweep rate $|dV_{dc}/dt|=10$ mV/s. Far from resonance, any piezomechanical response of the sample should be either in phase or out of phase by 180° with respect to the $V_{PFM}$. The in-phase PFM signal contains most of the essential information of a PFM measurement,[19] while the out-of-phase PFM signal is negligibly small. Based on **FIG. 3**(c), the *X*-output (in-phase) piezoresponse signal is correlated with the capacitance signal: as $V_{dc}$ decreases, the piezoresponse signal is suppressed and when $V_{dc}$ increases, the piezoresponse signal becomes enhanced. Both hysteresis loops (**FIG. 3**(b,c)) are stable as $V_{dc}$ swept forward and backward for successive cycles. Parametric plotting of the *X*-output piezoresponse versus capacitance (**FIG. 3**(d)), shows that the hysteresis near $V_{dc}=0$ V is highly correlated; however, there are significant deviations at both low and high values of the capacitance. Although the piezoresponse measurements are performed at a single location, this behavior is typical of measurements performed at other locations (see Appendix[20]).

### D. Quantitative PFM analysis

A non-zero piezoresponse implies that there is a lattice distortion as a result of the applied voltage. Therefore a quantitative analysis of the PFM[19, 21-23] measurement is performed (See Ref. 20 for a detailed description of the analysis). A series of piezoresponse spectra (**FIG. 4**(a)) is acquired with $V_{PFM}$=20 mV and $F_{PFM}$ from 285 kHz to 305 kHz under a sequence of bias conditions $V_{dc}$= 0V→-10V→10V→0V. The open circles in the intensity graph highlight the contact resonant frequency trajectory as $V_{dc}$ is modulated. For better illustration, the contact resonant frequency is plotted with respect to $V_{dc}$ in **FIG. 4**(b). The prominent resonant frequency change as well as the hysteretic behavior suggests a change in the height as well as the restoring force of the LaAlO$_3$ surface as applied $V_{dc}$ varies. **FIG. 4**(b) shares the same color legend as (a), which represents the PFM amplitude at each resonant frequency.

The bias-dependence of the effective piezoelectric strain tensor element $d_{33}^{eff}(V_{dc})$ (**FIG. 4**(c)) can be used to quantify the lattice distortion. This coefficient is obtained by fitting the piezoresponse spectra (**FIG. 4**(a)) to a simple harmonic oscillator (SHO) model[20] as a function of $V_{dc}$. The carrier density change at the interface (**FIG. 4**(d)) is calculated as follows: $\delta n = n(V_{dc}) - n(-10\text{V}) = (1/eA)\int_{-10\text{ V}}^{V_{dc}}(C(V) - C(-10\text{V}))dV$. By plotting parametrically $d_{33}^{eff}(V_{dc})$ versus $\delta n$, (**FIG. 4**(e)), a linear relationship between the strain tensor and carrier density change is revealed: $d_{33}^{eff} = d_{33}' \times \delta n + d_{33}''$, where $d_{33}' = 0.96$ ($10^{-12}$ pm cm$^2$/V) and $d_{33}'' = -0.63$ pm/V. In this manner, the effective piezoelectric strain tensor can be said to provide a direct measure of local carrier density.

The net vertical surface displacement as a function of $V_{dc}$ at the location of the AFM tip can be obtained by integrating $d_{33}^{eff}(V_{dc})$ with respect to the applied voltage $V_{dc}$:

$$\Delta z(V_{dc}) = \int_0^{V_{dc}} d_{33}^{eff}(V)\,dV$$ (**FIG. 5**(a)). The calculated surface displacement shows the lattice distortion under dc bias, which is due to both LAO polarization and the interface distortion. The displacement in both sides of the MIT (**FIG. 5**(a)) is qualitatively different: for negative $V_{dc}$, the displacement is linear, which means that there is a constant piezoresponse, plausibly from LAO polarization; for positive $V_{dc}$, the displacement is quadratic, which implies the distortion is associate with the charge accumulation at interface. The contribution from the LAO layer can be estimated by a linear fit in the insulating regime (**FIG. 5**(a)). The interface is obtained by subtracting the LAO contribution (**FIG. 5**(b)). The calculated dilation $\Delta d'(V_{dc}=5\text{V}) - \Delta d'(V_{dc}=-5\text{V}) = 0.11\,\text{Å}$ is consistent with previous experimental[24] and theoretical[25-27] reports.

### E. Simultaneous piezoresponse and capacitance measurement

To better understand the relationship between the capacitance and piezoresponse signals, the two signals are measured simultaneously with $V_{dc}$ abruptly changing into or out of the MIT region (**FIG. 6**). The piezoresponse is integrated over a frequency band centered on the contact resonance frequency (~270 kHz), while the capacitance is monitored in parallel according to the arrangement in **FIG. 3**(a). Strong correlations exist between the PFM and capacitance signals for each of the voltage ranges explored, which led further support to the thesis that PFM can be employed as a sensitive local probe of the gate-tuned carrier density.

### F. Time-resolved PFM measurements

A large capacitance enhancement on LAO/STO near the MIT was first reported by Li *et al.*[13]. A strong frequency dependence was observed, with enhanced capacitance values observed for frequencies below ~50-100 Hz. **FIG. 7**(a) shows *CV* measurements performed for Device A at frequencies ranging from 7 Hz – 100 Hz. An ac excitation voltages $V_C$=20 mV is employed, and $V_{dc}$ is varied from 1 V to -6 V in steps of 0.1V. **FIG. 7**(a) shows the result from device A. As the device is depleted of electrons ($V_{dc}$ decreasing from 0 V to -2 V), a capacitance upturn is observed. At its peak, the capacitance is enhanced significantly beyond the geometric limit, in accordance with prior reports[13]. The capacitance enhancement also shows a frequency dependence (**FIG. 7**(a)) that is in excellent agreement with Li et al (Ref. 13). At excitation frequencies above $F_c$~70 Hz, the capacitance enhancement decays, and vanishes completely by $F_c$=1 kHz. This enhancement is not present for positive gate sweeps (**FIG. 7**(b)); such sweeps were not reported in Ref. 13.

To gain insight into this phenomenon, time-resolved piezoresponse measurements are applied for different tip locations on the devices. The experimental setup is illustrated in **FIG. 3**(a): an ac excitation voltage $V_{PFM}$=20 mV at the resonant frequency together with a step-function pulse (blue curve, $V_{dc}$ switches between -2V and +2V) are applied to the LAO/STO interface. Time-resolved PFM measurements are performed using a lock-in amplifier with time constant $T_C = 1$ ms and frequency rolloff 6 dB/oct. As a reference, the lock-in amplifier response to a sudden amplitude-modulated test wave is shown, confirming that the bandwidth of the instrumentation exceeds that of the sample response by approximately one order of magnitude.

**FIG. 7**(c) and (d) show the time-resolved piezoresponse (note as *R(amp)*) as a $V_{dc}$=±2V pulse is turned on and off for Device A. Fitting the piezoresponse to an empirical expression $R(amp) = a_1 \times \tanh((t-t_0)/t_1) + a_2$ yields a characteristic response time $t_c = 1.76\, t_1$, defined as the full width at half maximum (FWHM) of *dR(amp)dt*, and corresponding frequency $1/t_c$. The response times in **FIG. 7**(c) and **FIG. 7**(d) are 11 ms and 12 ms respectively, corresponding to $1/t_c$ = 90 Hz and 84 Hz, respectively. These frequencies agree well with the observed frequency response in the *CV* experiments.

The time-resolved piezoresponse measurements are reproducible at different locations on the device. At each location, measurements are performed 5-10 times and the mean value is computed as well as the standard deviation. **FIG. 7**(e) and (f) show statistical results of measured response time on device A. The mean time for all these investigated locations is 9.3 ms in **FIG. 7**(e) and 11 ms in **FIG. 7**(f), which corresponds to ~108 Hz and 91 Hz respectively. Such results also agree with the frequency dependence of capacitance enhancement. Similar results are observed for the other devices (see Appendix).

### G. Spatially resolved PFM imaging

PFM has previously been reported to be highly correlated with the MIT[18]. Considering the observed linear relation between piezoresponse and carrier density **(FIG.** 4(e)**)**, PFM imaging can serve as a powerful local probe of interfacial charge density. Such spatially resolved PFM imaging can therefore provide insight into possible origins of hysteretic behavior and capacitance enhancement near the depletion region, which has not been reported before.

PFM imaging introduces some technical challenges. For example, if the AFM tip picks up a small particle or otherwise changes as it is scanning, the resonant frequency can shift. To minimize the influence of small frequency shifts during PFM imaging, a dual-ac resonance tracking (DART)[28] technique is used, with the parameter $f_{ac1}$ = 284.6 kHz , $f_{ac2}$ = 287.6 kHz and $V_{ac1} = V_{ac2} = 20$ mV. Each high-resolution image (128 lines) takes approximately 22 minutes to acquire.

FIG. 8 shows dual-frequency PFM amplitude images over a 500 nm × 500 nm area on the top-gate electrode of device A. Significant spatial variations in the critical voltage for strong enhancement of PFM on 25-50 nm scales are observed (**FIG. 8**(b-f)), while a superimposed inhomogeneity of several hundreds of nanometers is found on Device C (**FIG. 9**). **FIG. 8**(a) shows the topographic image and (b-f) are PFM images taken under decreasing bias $V_{dc}$= 3V, 2V, 1V, 0V, -2V, which traverse the MIT transition (**FIG. 3**(b)). Notice that for (b-f), each image is subtracted the mean value (from (b) to (f) is 4.24 a.u., 3.10 a.u., 1.91 a.u., 0.82 a.u., 0.35 a.u., respectively) to show a better PFM contrast. When $V_{dc} \geq 2$V, the carrier density at interface is relative high. PFM response is strong everywhere except for a few isolated regions. Also the inhomogeneity of the PFM signal is clearly shown in **FIG. 8**(b,c). In the images shown, green areas correspond to high PFM response (relative higher carrier density), while red regions have lower PFM response (relative lower carrier density). As $V_{dc}$ decreases in the range 2> $V_{dc}$ >-2 V, capacitance signal drops rapidly, indicating the depletion of mobile carriers at the interface. Within this range, the PFM signal becomes suppressed in some areas and PFM contrast becomes smaller (**FIG. 8**(d)-(e)), indicating large local variations in carrier density. When $V_{dc} \leq$-2V, the PFM signal becomes small with greatly suppressed spatial variations (**FIG. 8**(f)), corresponding to the interface globally switching to an insulating state. The evolution of PFM contrast from **FIG. 8**(b) to (f) shows direct evidence of ~30-50 nm-diameter conducting islands at the interface.

The observed fine spatial structures in PFM images are stable and reproducible, evidenced by the continuous scanning over the same area (**FIG. 9**) and scanning with fast and slow-scan axes reversed (**FIG. 10**). **FIG. 9** shows two sets of continuous PFM scanning under $V_{dc} = 3V$ ((a,b)) and $V_{dc} = 1.5V$ ((c,d)) separated by a time interval of 22 minutes. The fine spatial structure in PFM images at same $V_{dc}$ are reproducible, indicating the stability of conducting and insulating domains at interface. **FIG. 10**(a, b) are PFM images taken under $V_{dc}$=3V with switched fast/slow scan axis, which demonstrate the fine structures in PFM image are reproducible regardless of the scan direction.

### G. Normalized root-mean-square (RMS) analysis

The PFM images reveal inhomogeneity of the carrier density at the interface. To check the inhomogeneity of the piezoresponse signal over the scanned areas with respect to different $V_{dc}$, a normalized root-mean-square (RMS) analysis is applied to PFM images (500×500 nm$^2$) taken under various values $V_{dc}$ on different devices. This analysis is repeated at several locations using different sets of PFM images. Before calculating the RMS value, each PFM image is averaged over 25 nm × 25 nm area to reduce contributions due to intrinsic noise. The RMS PFM signal is then calculated as follows:

$$Normalized\ RMS = \frac{RMS(V_{dc})}{\overline{x(V_I)}} = \frac{\sqrt{\frac{1}{n}\sum_{i=0}^{n-1}|x_i(V_{dc})|^2}}{\frac{1}{n}\sum_{i=0}^{n-1}x_i(V_I)}$$

Normalization is taken with respect to the average signal in the insulating region ($V_I$=-5V). With the resolution of 128×128 for all PFM images, $n$=16384 in the formula. The normalized RMS signals for device A, B and C are illustrated in **FIG. 11**(a-c), respectively**.** All of the normalized RMS curves show similar behavior: the normalized RMS signals are enhanced as $V_{dc}$ increases in the conducting region, indicating large inhomogeneity at the interface. This statistical result is consistent with the observed PFM contrast change in **FIG. 8**.

## IV. DISCUSSION

The combined PFM and capacitance measurements performed on top-gated LAO/STO devices reveal strong correlations between structural, capacitive, and metallic properties. Bark *et al.*[18] proposed that the electrically-induced oxygen vacancy migration in the LAO layer is responsible for the hysteretic PFM response. The PFM is performed at relatively high frequencies (around 290 kHz), whereas oxygen vacancies have too small of a mobility $\sim 10^{-13}$ cm$^2$/Vs, [29] for the piezoresponse signal to be attributed to oxygen vacancy motion. Furthermore, it is observed that in air the piezoresponse hysteresis exists even under a complete lack of local driving, but vanishes in vacuum with the same experiment setup[30], which cannot be explained by oxygen vacancy mechanism. Finally, the samples described here are annealed in partial oxygen pressure ambient at 600 °C, conditions under which the majority of oxygen vacancies are expected to be removed in the STO substrate[31].

The electromechanical response can be attributed to Jahn-Teller effects [32,33], which proposes the electron accumulation on Ti-3d band at the LAO/STO interface can induce oxygen octahedra distortion. There is evidence from both transmission electron microscopy[24] and X-ray[34] measurements that when LAO is grown on (001) STO, distortions in TiO$_6$ octahedra happen at the interface. Such lattice distortions at the interface result in a biaxial strain and a strain gradient that can induce ferroelectric[35] and flexoelectric[36] polarization in STO near the interface. Theoretical calculations[25-27] of LAO/STO system also reveal carrier-mediated distortions at the interface. The calculated dilation at the interface is about $\delta z = 0.15$ Å is consistent with what is reported here.

The Jahn-Teller mechanism also helps to explain the strong correlation between capacitance signal and piezoresponse (**FIG. 3**, **FIG. 6**). In the insulating phase (large negative $V_{dc}$), the capacitance is small and approximately bias-independent. The piezoresponse is similarly small and independent of $V_{dc}$. As $V_{dc}$ increases, electrons will eventually be introduced to the interface, leading to an enhancement of capacitance signal. These electrons can occupy/evacuate the Ti-3d band at interface under $V_{PFM}$ modulation, which gives rise to a stronger oxygen octahedra distortion as well as an enhanced piezoresponse signal that scales with the local carrier density.

The spatially resolved PFM images show large local variations in PFM contrast, suggesting that the MIT is strongly inhomogeneous. Previous Kelvin probe force microscopy studies[37, 38] also show an inhomogeneous distribution of the surface potential in LAO/STO samples, which suggests an inhomogeneous carrier density at the interface. The frequency response of the PFM and CV measurements is marked by charging-discharging dynamics of nanoscale islands observed in PFM images. At the MIT, the characteristic frequency response compares well to the conductance measurement for a randomly mixed conductor-insulator system near the percolation threshold[39]. The observed frequency dependence of capacitance enhancement may be understood in terms of an intrinsic time scale for island charging-discharging dynamics.

To summarize, capacitance characterization and PFM are combined to help understand the origin of observed piezoresponse signal, and the mechanism for hysteretic behavior in CV and PFM measurements. The frequency dependence of the capacitance enhancement in LAO/STO matches that of local PFM measurements and is otherwise highly correlated. A quantative analysis of the piezoresponse indicates there is a structural distortion associated with the gate-tuned MIT at the LAO/STO interface. The carrier-mediated structural distortion is understood as a Jahn-Teller effect, with a magnitude that agrees with theoretical calculations. Spatially resolved PFM provides a direct visualization of the island charging/discharging process at interface under gate tuning. Our experiments provide a fuller understanding of the interplay between electrons and lattice degrees of freedom, and may help to find more potential applications of oxide-based nanoelectronics.

## ACKNOWLEDGEMENTS

This work was supported by NSF DMR-1104191 (JL) and DMR-1124131 (CBE, JL). The work at University of Wisconsin-Madison was supported by the NSF under Grant No. DMR-1234096 and AFOSR under Grant No. FA9550-12-1-0342 .

**FIGURE LEGENDS**

**FIG. 1** (Color online) (a) Photograph of capacitor devices fabricated on a LAO/STO sample with leads attached. The STO substrate is square with a size 5 mm x 5mm. Sample is mounted on a ceramic chip carrier. (b) I-V measurement between two separated electrodes that directly contact to the interface. The linear I-V behavior shows the electrode-interface contact is ohmic. The conductance of interface between these two electrodes is $3.2 \times 10^{-5}$ S based on the linear fit of IV curve, proving the interface is in conducting phase. Results are shown for sample A.

**FIG. 2** (Color online) Resistance measurements between two electrodes that contact the interface (a) Schematic diagram showing the interface resistance measurements (b) Resistance of the interface as a function of top gate bias. Results are shown for device A.

**FIG. 3** (Color online) Simultaneous PFM and capacitance measurement (a) Sample layout with circuit diagram of PFM measurement (red) is combined with a capacitance bridge (green). (b) $C$-$V_{dc}$ curve taken at room temperature at $F_C$=1 kHz with the sweep rate of $V_{dc}$: 1/15 V/s, showing a hysteretic metal to insulator transition as $V_{dc}$ is varied. Piezoresponse in-phase (c) signals at different values of $V_{dc}$ with the excitation voltage given by $V_{PFM}$=20 mV, $F_{PFM}$=293 kHz. (d) Plot of piezoresponse versus capacitance signal shows a positive correlation between them. Results are shown for device A.

**FIG. 4** (Color online) Quantitative analysis of PFM spectrums on device A. (a) Piezoresponse within a frequency band from 285 kHz to 305 kHz under various dc bias conditions ( $V_{dc}$: 0 V→-10 V→10 V→0 V, step 0.2 V). The open circles in this intensity graph represent the contact resonant frequency peak under each dc bias. (b) Contact resonant frequency is extracted from (a) and plotted as a function of $V_{dc}$. The color of each data point represents the piezoresponse

amplitude at each resonant frequency. (a) and (b) share the same color scale legend. (c) $d_{33}^{eff}(V_{dc})$ is acquired from piezoresponse spectra in (a) using SHO model fitting. (d) Carrier density change: $\delta n$ at the interface as a function of $V_{dc}$. $\delta n$ is calculated from equation:

$$\delta n = n(V_{dc}) - n(-10\,\text{V}) == \frac{\int_{V=0}^{V_{dc}}(C(V)-C(-10\text{V}))dV}{e \times \pi r^2} \quad \text{with } r = 50\,\mu\text{m}.$$

**FIG. 5** (Color online) (a) Surface displacement calculation based on $d_{33}^{eff}(V_{dc})$ in **Fig.4** (c) from equation $\Delta z(V_{dc}) = \int_0^{V_{dc}} d_{33}^{eff}(V)dV$. At $V_{dc}<0$ region, the displacement is linear with respect to the $V_{dc}$. The linear response is fitted as black dashed line. (b) $\Delta d'$ as a function of $V_{dc}$ acquired from (a) subtract the linear fitting, which we interpret as interface distortion.

**FIG. 6** (Color online) Simultaneous time-dependent PFM and capacitance at fixed top gate biases. (a) $V_{dc}$=-5 V. (b) $V_{dc}$=-1 V. (c) $V_{dc}$=2 V. Top graph shows the $V_{dc}$ bias history. $V_C$ and $V_b$ are applied at the capacitance bridge frequency 14Hz. PFM response is measured at a frequency band centered around the contact resonance frequency between 290 kHz to 310 kHz with 200 Hz steps. To avoid the influence of resonance frequency changes, the PFM response is integrated in the frequency band from 290 kHz to 310 kHz. Results are shown for device B.

**FIG. 7** (Color online) Capacitance enhancement and time resolved PFM analysis. Capacitance is measured as dc bias decreases (a) or increases (b). Curves are taken at room temperature with frequency $F_C$=7Hz to 100Hz. (c) Time resolved piezoresponse during an applied step-function dc bias. (d) Piezoresponse as the $V_{dc}$ bias is removed. Red circles are data and the green line shows a fit to an empirical expression $R(amp) = a_1 \times \tanh((t-t_0)/t_1) + a_2$. The response time $t_c$ is defined by FWHM of $dR(amp)/dt$ and corresponds to frequency $f_c=1/t_c$. The purple squares show

the lock-in amplifier equipment response time is around 1 ms ($TC$=1 ms, rolloff 6 dB/oct) from fitting. (e), (f) are time resolved piezoresponse analysis at different locations. At each location, measurements are repeated several times and the mean $t_c$ and standard deviation $\sigma$ are calculated. The mean value for all $t_c(local)$ is also noted, as $t_c$ in the graph. Data shown is from device A.

**FIG. 8** (Color online) Spatially-resolved dual frequency PFM images show inhomogeneity at interface. The height image (a) and corresponding PFM images (b-f) are shown at bias values $V_{dc}$ = (b) 3V, (c) 2V, (d) 1V, (e) 0V (f) -2V with the mean value 4.24 a.u., 3.10 a.u., 1.91 a.u., 0.82 a.u., 0.35 a.u. subtracted. Scan size 500 nm×500 nm; results shown are from device A.

**FIG. 9** (Color online) Spatially resolved dual frequency PFM images (500nm×500nm) on device C. Images are taken under $V_{dc}$: 1.5 V (a),(b) and 2.5 V (c),(d) at different time after the bias application. The average PFM signals for (a) 0.22 a.u., (b) 0.23 a.u., (c) 0.33 a.u., (d) 0.34 a.u. are already subtracted from each image for the purpose of showing a better contrast.

**FIG. 10** (Color online) Dual-frequency PFM images (500 nm×500 nm) on device C. Images are taken under $V_{dc}$=3 V with scan angle 0° (a) and 90° (b) to show the fine structures in PFM image are reproducible regardless of the scan direction. Image with scan angle 0° is rotated 90° counter-clockwise in (a) for easier comparison with (b). The blue dash line circled areas are examples to show the reproducibility of fine structures.

**FIG. 11** (Color online) (a-c) shows normalized RMS analysis for PFM images acquired over different locations on Device A, B and C respectively.

FIGURES

(a) 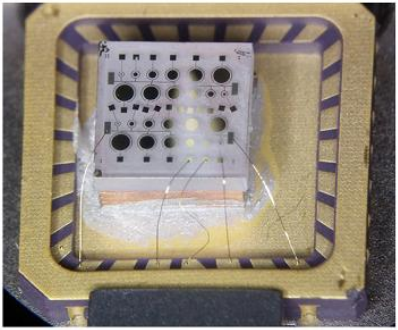 (b) 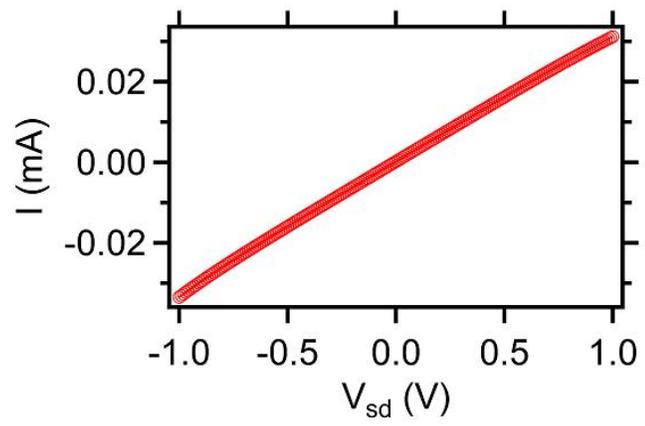

**FIG. 1**

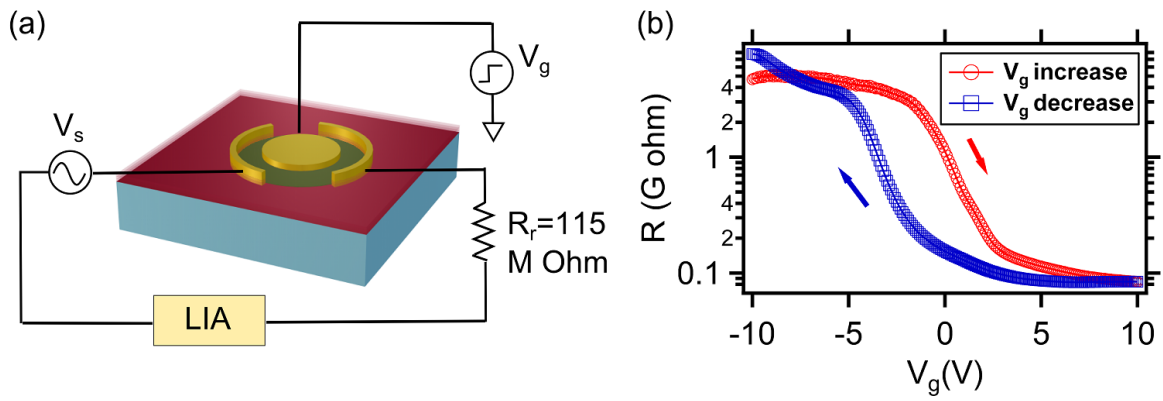

**FIG. 2**

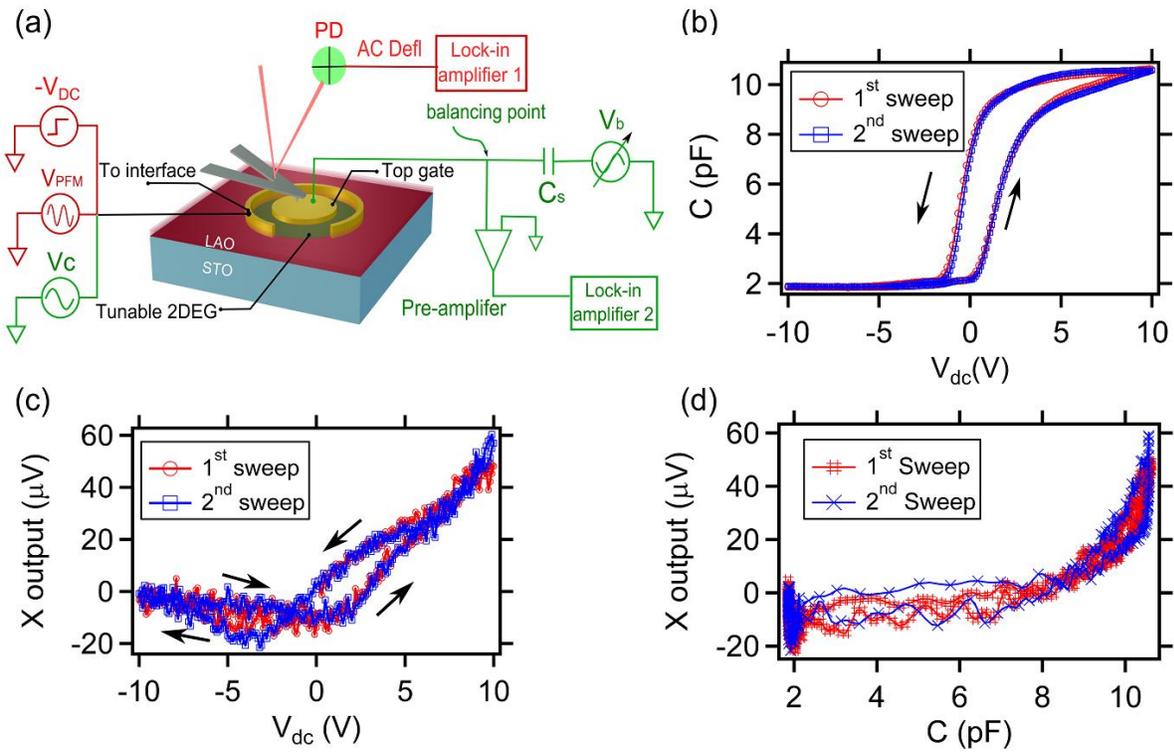

**FIG. 3**

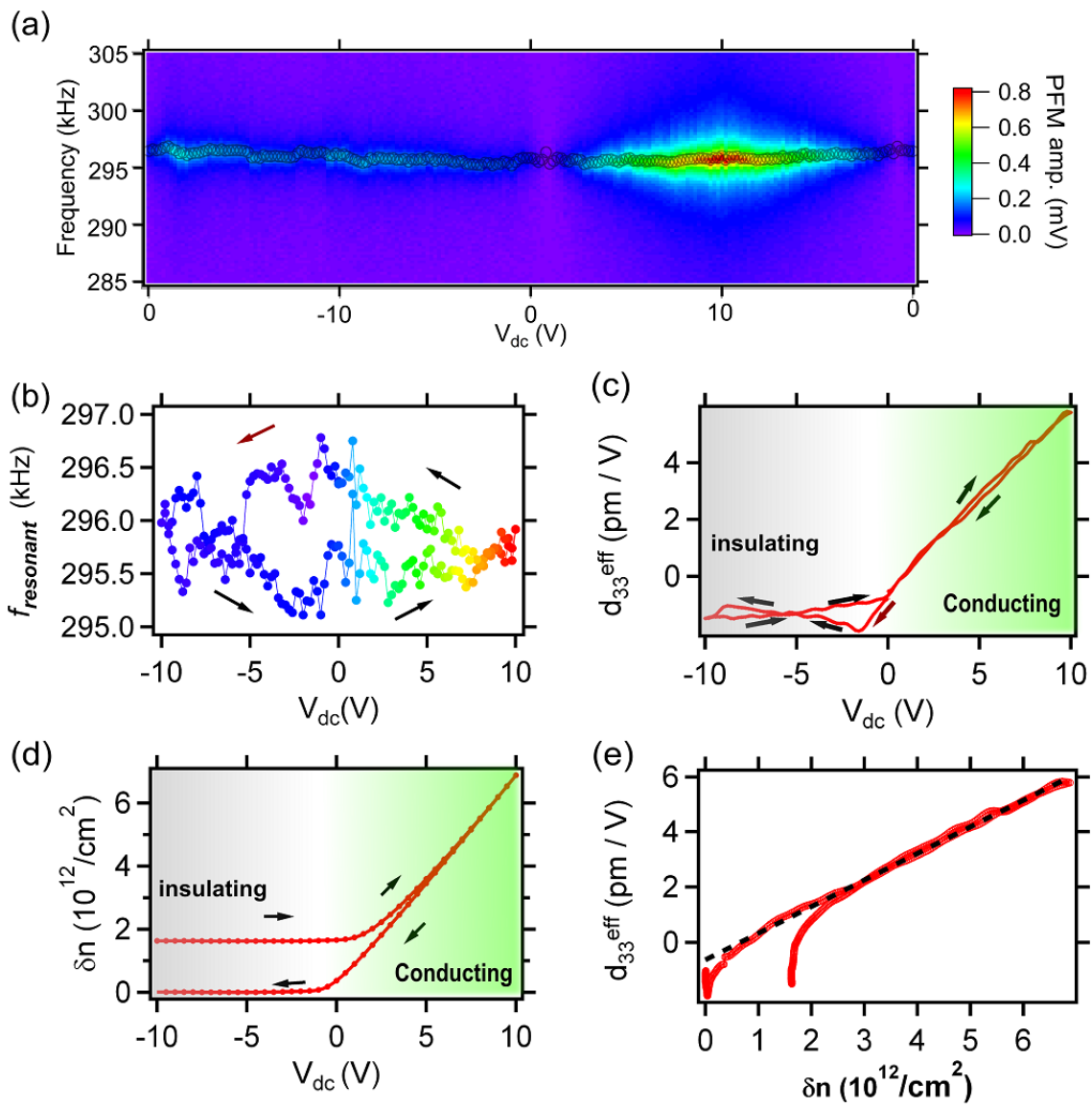

**FIG. 4**

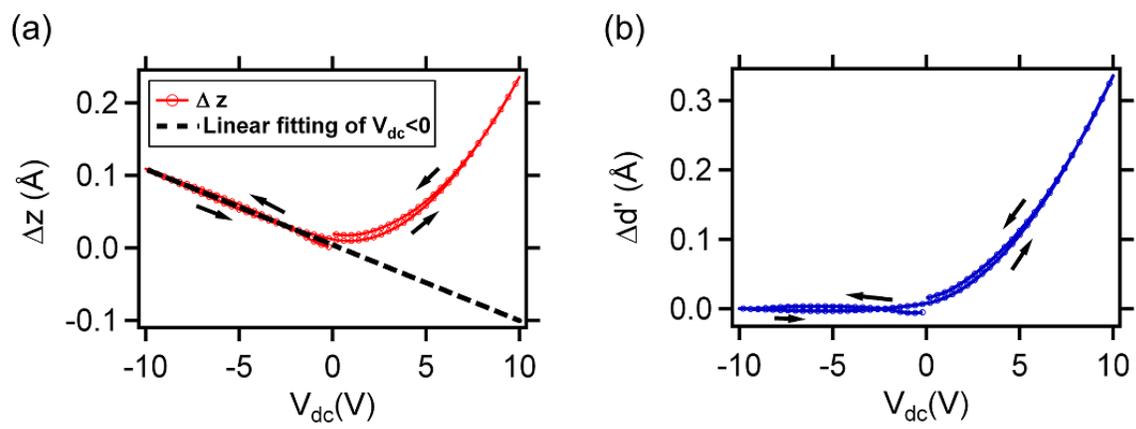

**FIG. 5**

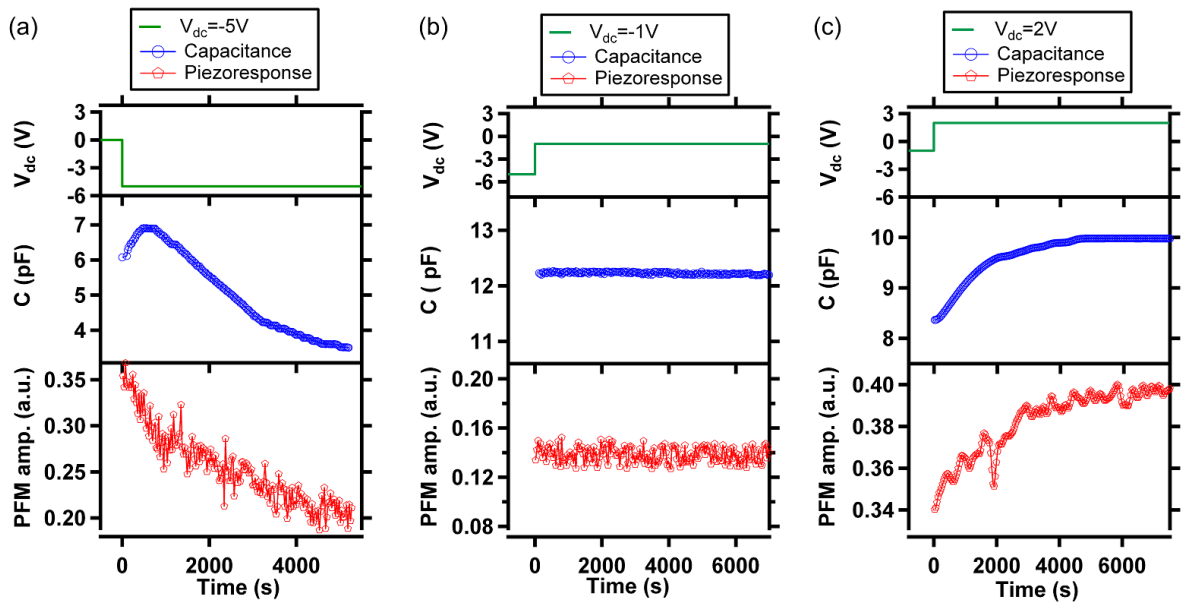

**FIG. 6**

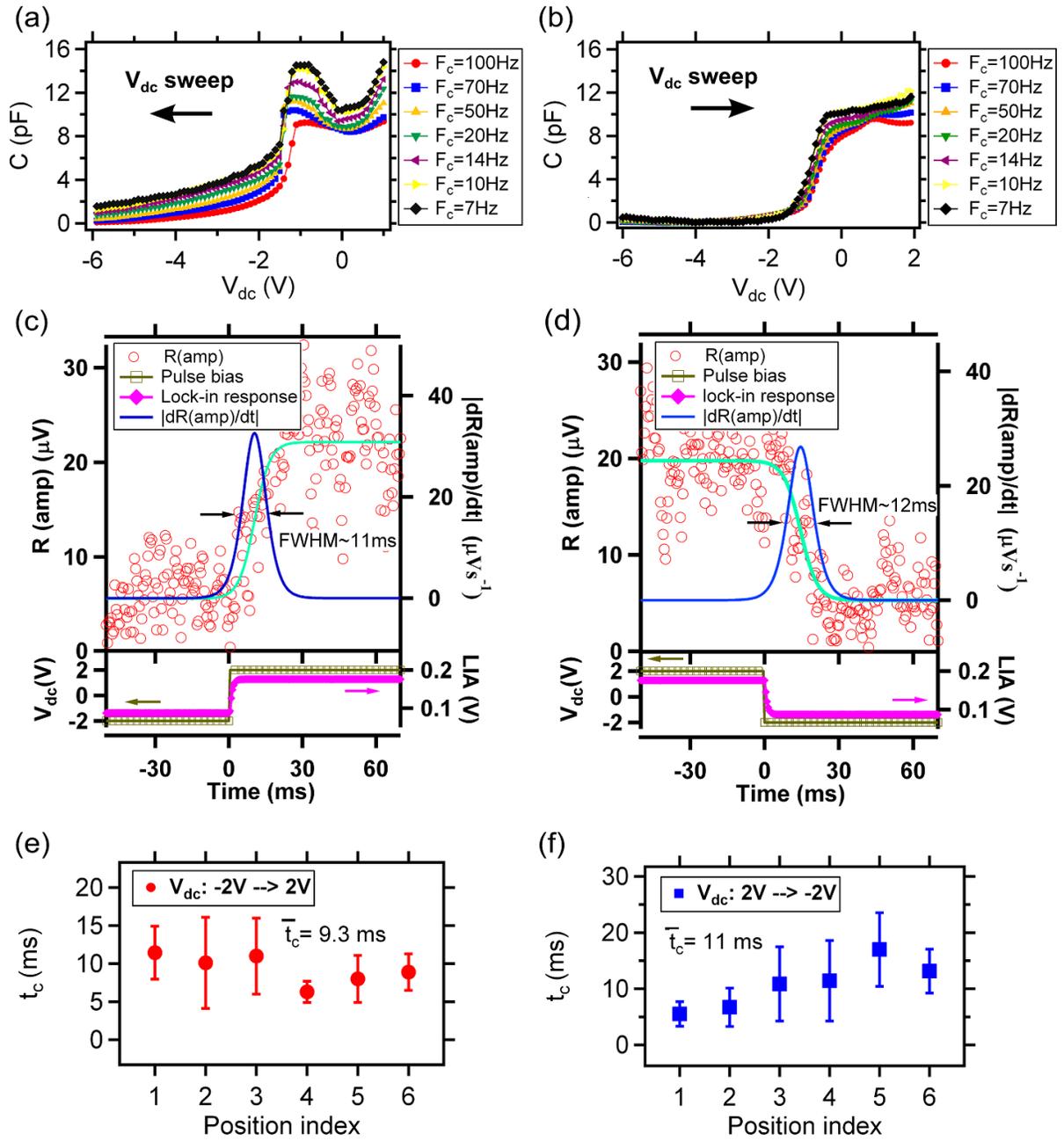

**FIG. 7**

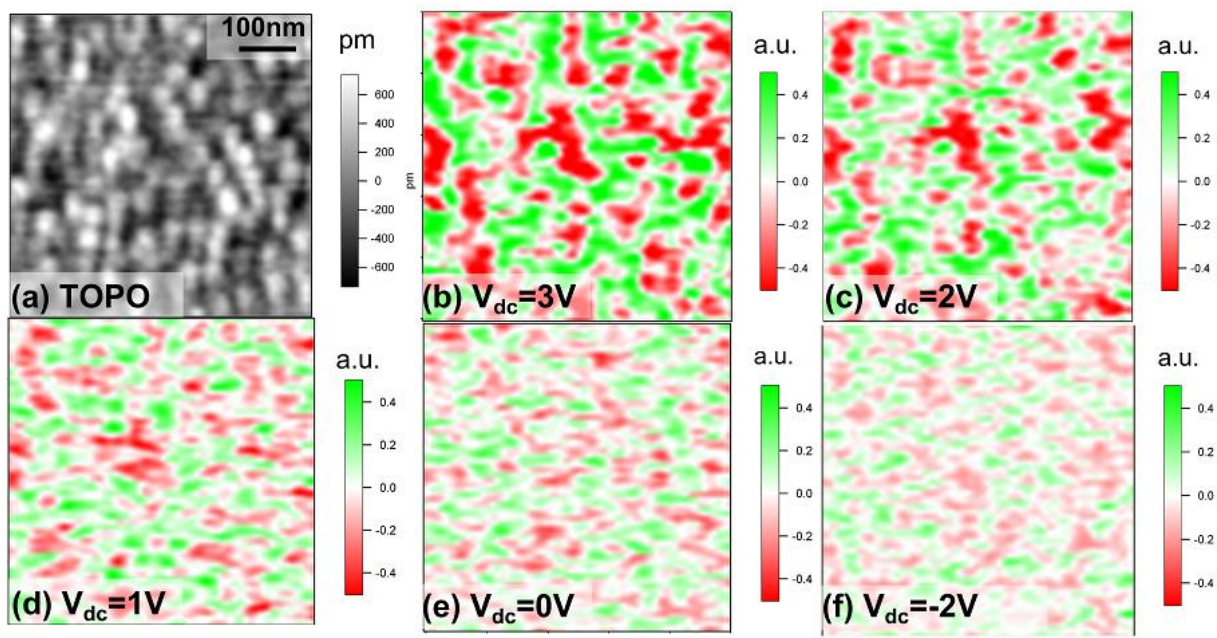

**FIG. 8**

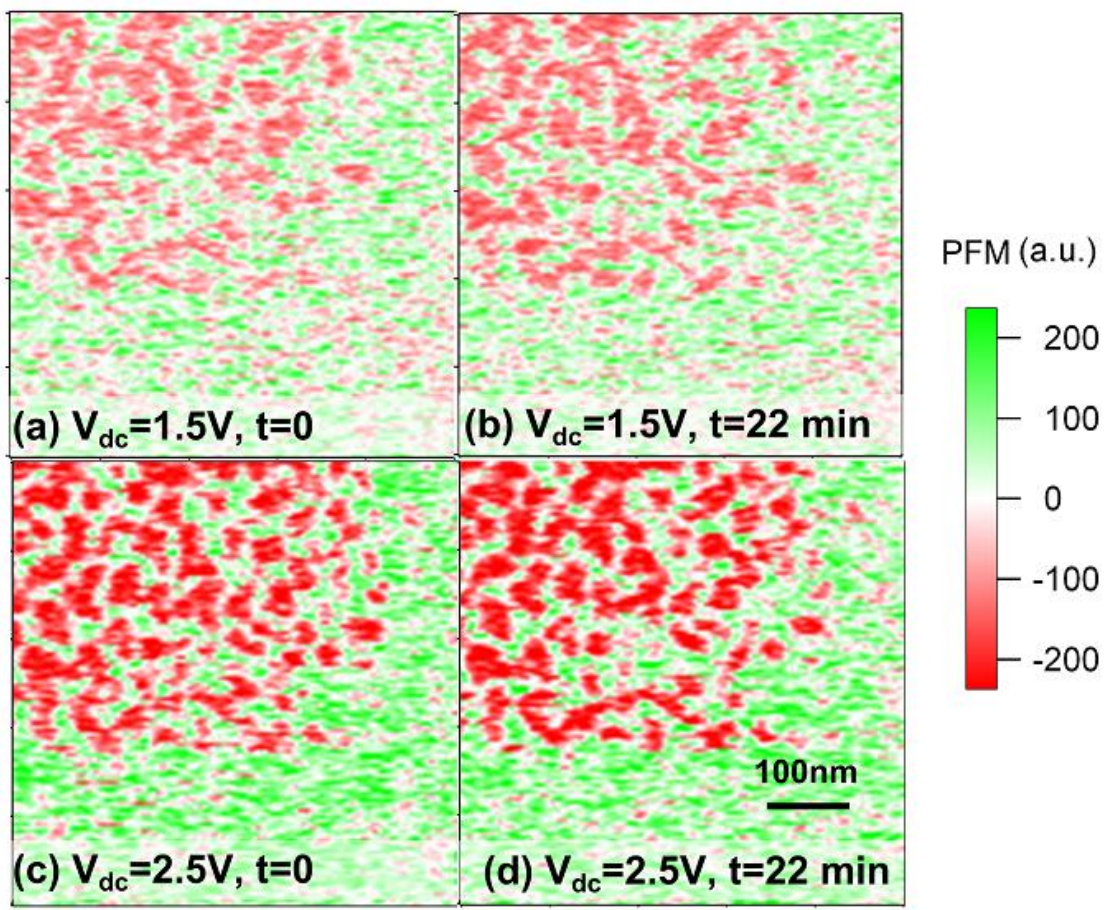

**FIG. 9**

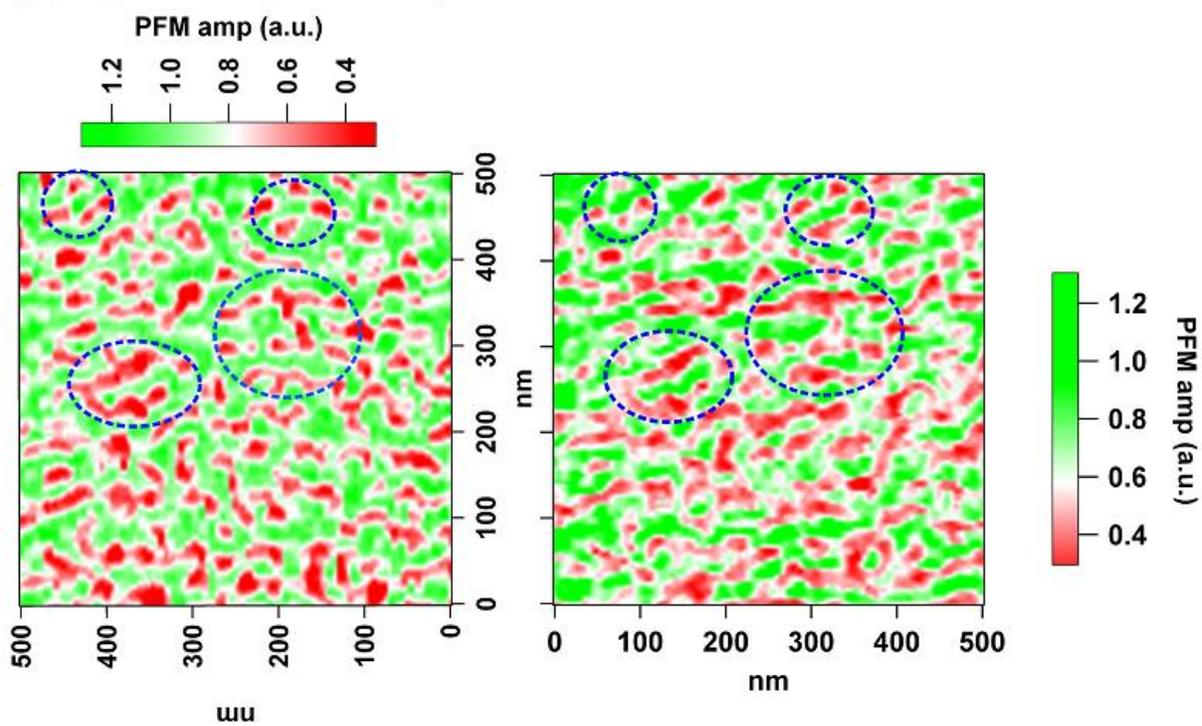

**FIG. 10**

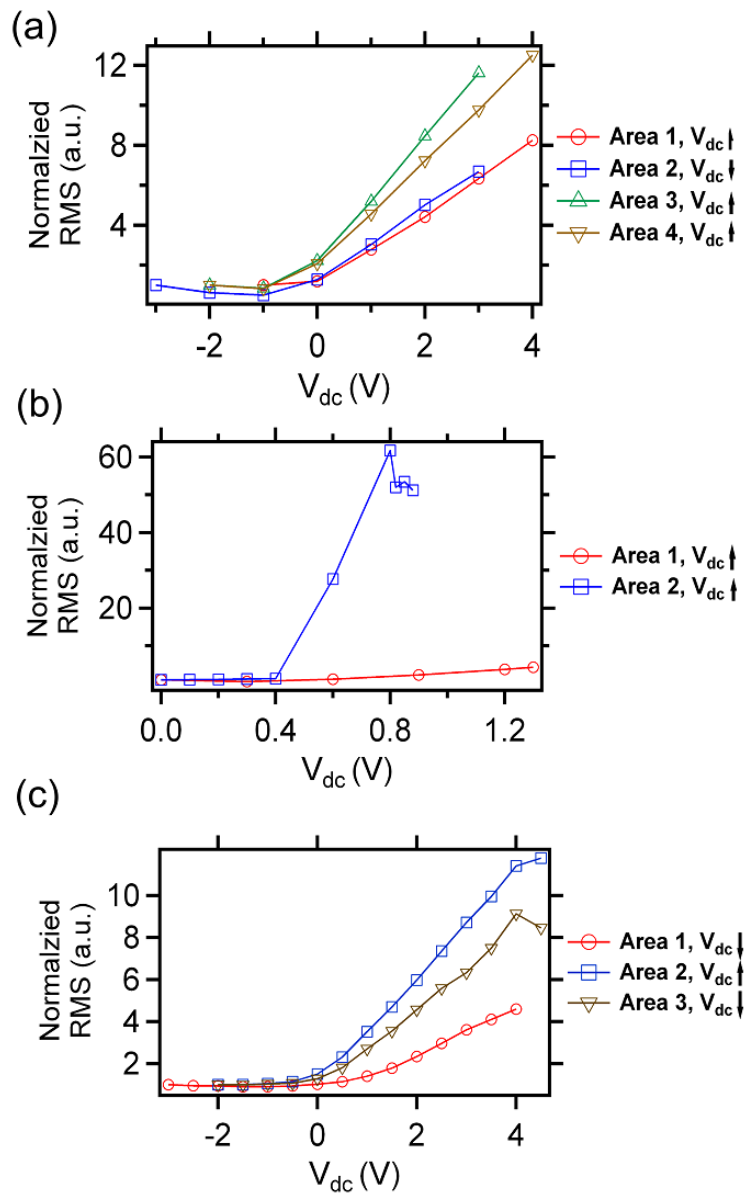

**FIG. 11**

# Appendix

Feng Bi,[1] Mengchen Huang,[1] Chung Wung Bark,[2] Sangwoo Ryu,[2] Sanghan Lee,[2] Chang-Beom Eom,[2] Patrick Irvin[1] and Jeremy Levy[1,*]

[1]*Dept. of Physics & Astronomy, University of Pittsburgh, Pittsburgh, Pennsylvania, 15260, USA*
[2]*Dept. of Materials Science, University of Wisconsin-Madison, Madison, Wisconsin, 53706, USA*

**Quantitative analysis of PFM measurement on device A:**

Piezoresponse Force Microscopy is based on the detection of electrical bias-induced surface deformation. The excitation voltage: $V(t) = -V_{dc} - V_{PFM}\cos(2\pi F_{PFM} t)$ is applied to the LAO/STO interface with the top electrode held at virtual ground. The surface deformation will be:

$$z = z_{dc} + A_z(F_{PFM}, V_{PFM}, V_{dc}) \times \cos(2\pi F_{PFM} t + \varphi_0) = \int_0^{V_{dc}} d_{33}^{eff}(V_{dc}) \times dV + d_{33}^{eff}(V_{dc}) \times V_{PFM} \times \cos(2\pi F_{PFM} t) \quad (2).$$

Here $A_z(F_{PFM}, V_{PFM}, V_{dc})$ is surface deformation amplitude under bias modulation, $d_{33}^{eff}$ is the effective strain tensor component that directly couples into the vertical motion of the cantilever and $\varphi_0 = 0$ or $\pi$, depending on the sign of $d_{33}^{eff}(V_{dc})$. The surface deformation couples to the cantilever displacement, which results in a change of deflection that is monitored with a lock-in amplifier. Due to the contact resonant enhancement, the surface deformation will be amplified by $Q$ times, the quality factor of the tip-surface contact resonant system. If the lock-in X-output is denoted by $X(\omega_0, V_{dc})$ and $\alpha$ as the sensitivity of AFM, then

$$d_{33}^{eff}(V_{dc}) V_{PFM} Q = X(\omega_0, V_{dc}) \alpha \quad (3).$$

. The sensitivity $\alpha = 67.75$ nm/V can be obtained by measuring the AFM force curve. The tensor component $d_{33}^{eff}(V_{DC})$ is calculated as follows:

$$d_{33}^{eff}(V_{dc}) = \frac{X(\omega_0, V_{dc})\alpha}{QV_{PFM}} \quad (4).$$

The surface displacement at the location of the AFM tip can be obtained by integrating with respect to the applied voltage

$$z_{dc}(V_{dc}) - z_{dc}(0\text{ V}) = \int_{0\text{ V}}^{V_{dc}} d_{33}^{eff}(V_{dc})dV \quad (5).$$

The surface displacement is calculated under different $V_{DC}$ based on the high frequency PFM measurement on device A. PFM spectral measurements (from 285 kHz to 305 kHz) are performed under different $V_{DC}$ modulated from 0V→-10V→10V→0V with step size 0.2V (**FIG. 4**(a)). During the measurement, the ac excitation voltage is given by $V_{PFM}$=20 mV.

Each spectrum is then fitted to a simple harmonic oscillator (SHO) model:

$$A(\omega) = \frac{A^{max}\omega_0^2}{\sqrt{(\omega^2 - \omega_0^2)^2 + (\omega\omega_0/Q)^2}} \quad (6).$$

where $\omega_0$ is the resonant frequency, $A^{max}$ is the amplitude and $Q$ is the quality factor. With $\omega_0$ and $Q$ at each dc bias, one obtains:

$$X'(V_{dc}) = \frac{X(\omega_0, V_{dc})}{Q} \quad (7).$$

Equation (4) and (5) can be simplified to

$$d_{33}^{eff}(V_{dc}) = \frac{X'(V_{dc}) \times 67.75 \,\text{nm/V}}{V_{PFM}} \tag{8}$$

$$z_{dc}(V_{dc}) = \int_0^{V_{DC}} \frac{X'(V_{dc}) \times 67.75 \,\text{nm/V}}{V_{PFM}} \times dV \tag{9}$$

Eq. (8) and Eq. (9) give the result of $d_{33}^{eff}(V_{dc})$ and surface displacement $z_{dc}(V_{dc})$, which are plotted in **FIG. 4**(c) and **FIG. 5**(a) separately. **FIG. 5**(a) shows the quantitative distortion within LAO/STO during the interface MIT.

**System response time measurement:**

To measure the lock-in amplifier response time, a function generator produces an amplitude modulated test wave (271 kHz sinusoid wave modulated by a square waveform with modulation rate: 1Hz and modulation depth: 50%). Such test wave is then measured by the lock-in amplifier with $TC$=1 ms, rolloff 6 dB/oct and reference frequency 271 kHz. The system response time is fitted to be 1ms, confirming that the bandwidth of the instrumentation exceeds that of the sample response by approximately one order of magnitude.

**Local piezoresponse measurements on different locations and samples:**

Local piezoresponse measurements in the main text show a counterclockwise hysteretic behavior. The in-phase piezoresponse signal gets suppressed as $V_{dc}$ decrease and enhanced as $V_{dc}$ increase. These features are typical for all the local piezoresponse measurement results taken at different tip locations and different devices. For example, **FIG. S1**(a)-(b) show the local piezoresponse measurements at a different tip position on device A with the excitation voltage given by $V_{PFM}$=50 mV, $F_{PFM}$=293 kHz. **FIG. S1**(c)-(d) are the same measurements applied to device B with $V_{PFM}$=50 mV, $F_{PFM}$=301 kHz. (The PFM excitation frequency is adjusted a little bit according to the small change of the tip-surface contact resonant frequency.) All these results in **FIG. S1** show similar behaviors as that in the main text.

**Time resolved piezoresponse measurements on device B:**

The time resolved piezoresponse results do not show clear device dependence, based on our measurements on different devices. Here **FIG. S2** shows the time resolved piezoresponse during an applied step-function of dc bias from -2V to 2V and from 2V to -2V. The response time is fitted to be 14 ms and 15 ms, which corresponds to 71 Hz and 67 Hz, respectively. Such results are close to that on device A. The results suggest the charging/discharging time scale is similar for these devices regardless of LAO thickness and growth conditions. This is probably due to the fixed distance (50 μm) between top gate and electrode to interface for all the devices.

**Spatially-resolved PFM images under varied $V_{dc}$ on devices B and C:**

2D PFM imaging is also performed on different devices. **FIG. S3** and **FIG. S4** are PFM amplitude images on device C and device B with the excitation voltage $V_{ac1} = V_{ac2} = 20$ mV at $f_{ac1} = 284.6$ kHz and $f_{ac2} = 287.6$ kHz. Both **FIG. S3** and **FIG. S4** shows island patches, which demonstrate the inhomogeneity at interface. These island patches, regions of high and low carrier density, evolve with changing dc bias. In the conductive phase, the piezoresponse amplitude becomes greatly enhanced and increasingly inhomogeneous.

There is some correlation between the height and PFM signals in the **FIG. S3**. These correlateions may arise from interactions between the Au top gate and interface, which is discussed in the **Ref**. 1.

**Spatially-resolved PFM images taken at different time frame on device B**

Under fixed dc bias, 2D PFM images are taken at different time frame on device B. With $V_{dc}$=0.6V, **FIG. S5**(a),(b) are taken at a time interval of 9 minutes. **FIG. S5**(c),(d) are also taken at the same time interval but with $V_{dc}$=0.8V. Results show that PFM signals evolve as time elapse. By comparing (c) and (d), the island patches can be seen to expand or shrink. The evolution of PFM images over time on device B agrees with the time scale of **FIG. 6** in the main text.

**Leakage current measurement:**

The *IV* measurements are performed between top gate and interface on all three devices. $V_{dc}$ is applied to top gate and the dc current is measured from electrode contacting the interface with a current-voltage amplifier. Samples are kept in dark overnight before measurement. Measurement result shows junction behavior.

# FIGURE LEGENDS

**FIG. S1** (a,b) Piezoresponse signals on device A at a different tip location from that in the main text with the excitation voltage given by $V_{PFM}$=50 mV, $F_{PFM}$=293 kHz under varied $V_{dc}$. (c,d) Piezoresponse measurements on Device B with the excitation voltage given by $V_{PFM}$=50 mV, $F_{PFM}$=301 kHz. *X*, *Y* output represent the in-phase and out-phase piezoresponse signal respectively.

**FIG. S2** (a) Time-resolved piezoresponse during an applied step-function dc bias from -2 V to 2 V. (b) Piezoresponse as the $V_{dc}$ bias switches back to -2 V. Red circles are data and the green line shows a fit to an empirical expression $R(PFM) = a_1 \times \tanh((t-t_0)/t_1) + a_2$. Blue curve (d(*R(PFM)*)/*dt*) is the derivative of green line with respect to *t*. Results are shown for device B.

**FIG. S3** Spatially-resolved PFM imaging (500 nm×500 nm) on device C over the same area in a relative large dc bias range (b) 0 V, (c) 1 V, (d) 2 V, (e) 3 V. The topography image is shown in (a). The PFM difference between (e) and (d) is plotted as (f).

**FIG. S4** Spatially-resolved PFM imaging (400 nm×400 nm) on device B over the same area in fine dc bias range (b) 0.4 V, (c) 0.6 V, (d) 0.8 V, (e) 0.85 V and (f) 0.88 V. Panel (a) is the corresponding surface topography for (b)-(e).

**FIG. S5** Spatially-resolved PFM images (400 nm×400 nm) on device B at different time frame with the dc bias to be (a)-(b) 0.6 V and (c)-(d) 0.8 V.

**FIG. S6** Leakage current measurements on all three devices between top gate and interface. $V_{dc}$ is applied to top gate, swept from 10 V to -10 V with rate 0.1 V/s.

**FIG. S7** (a) Histogram of PFM images **FIG.8 (c)-(f)** in the main text. (b) Simple calculation of relative conducting area/insulating area based on (a). (c) Corresponding capacitance value under each dc bias. Result is shown for device A.

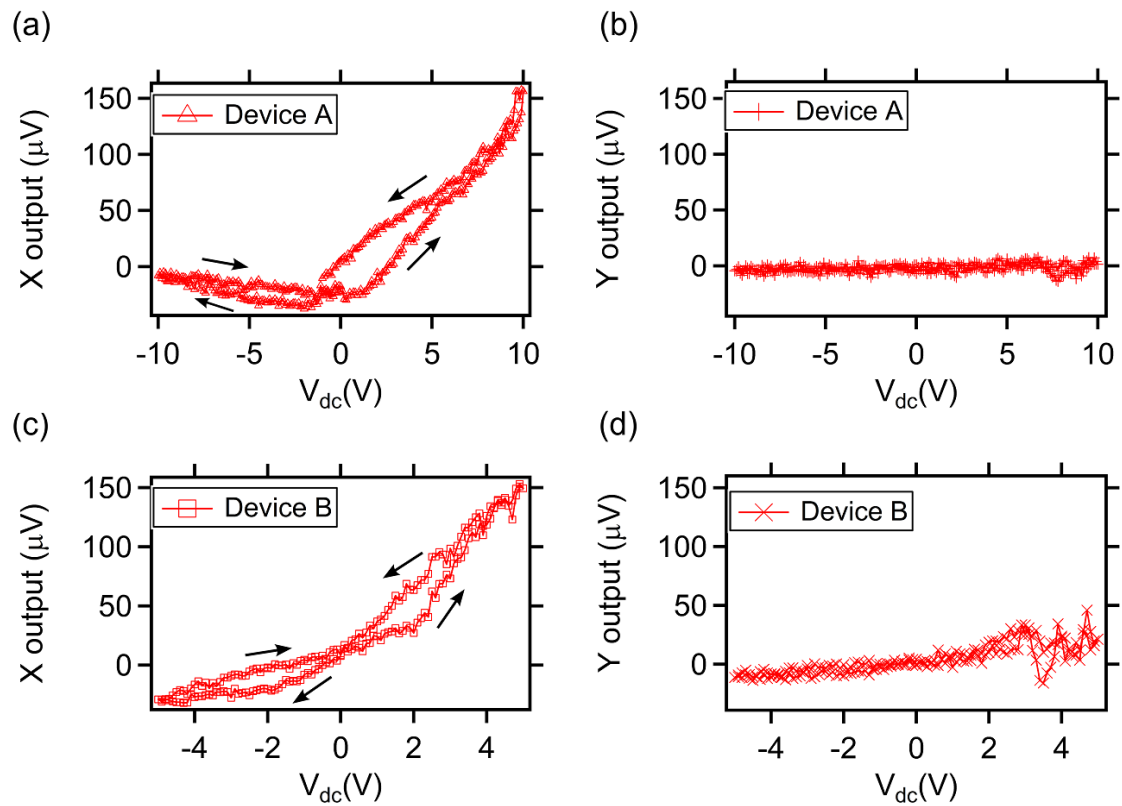

**FIG.S1**

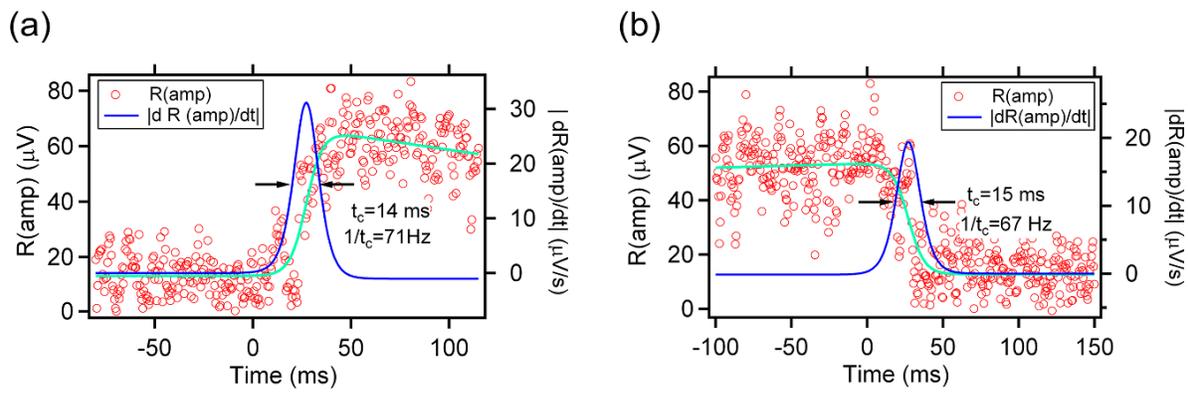

**FIG.S2**

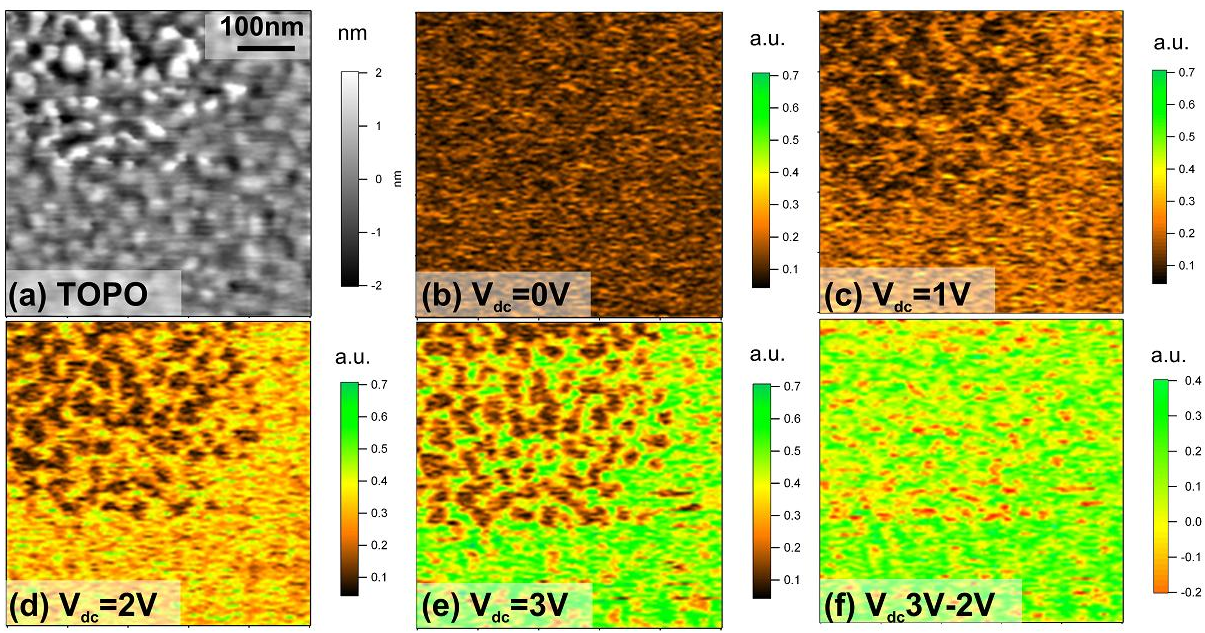

**FIG.S3**

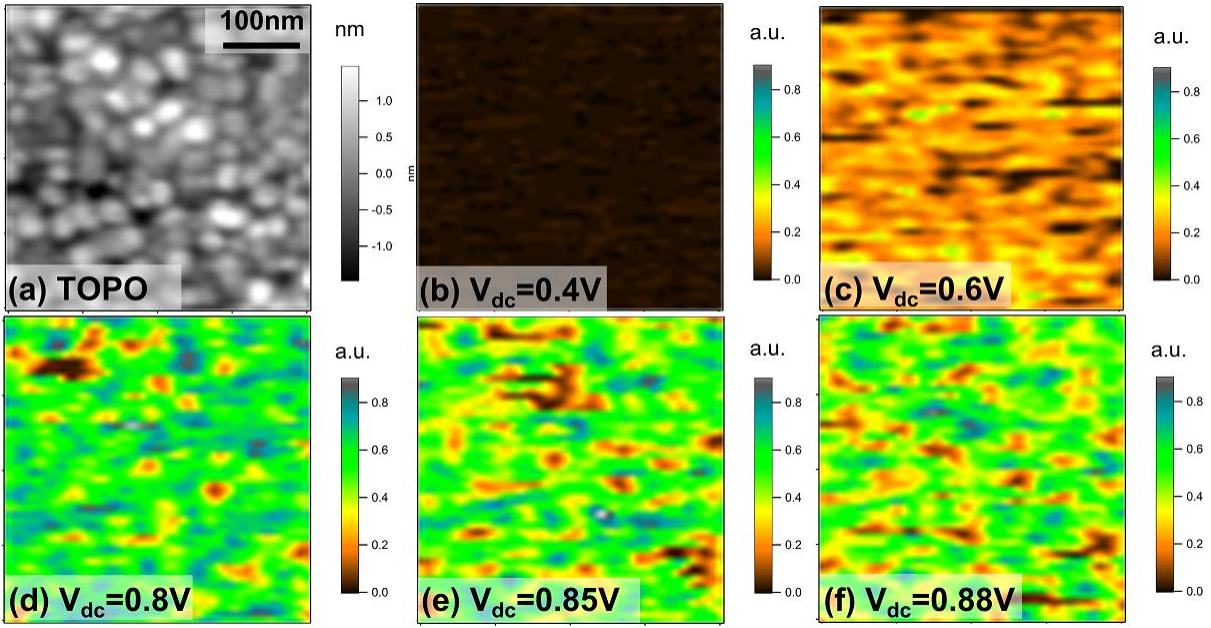

**FIG.S4**

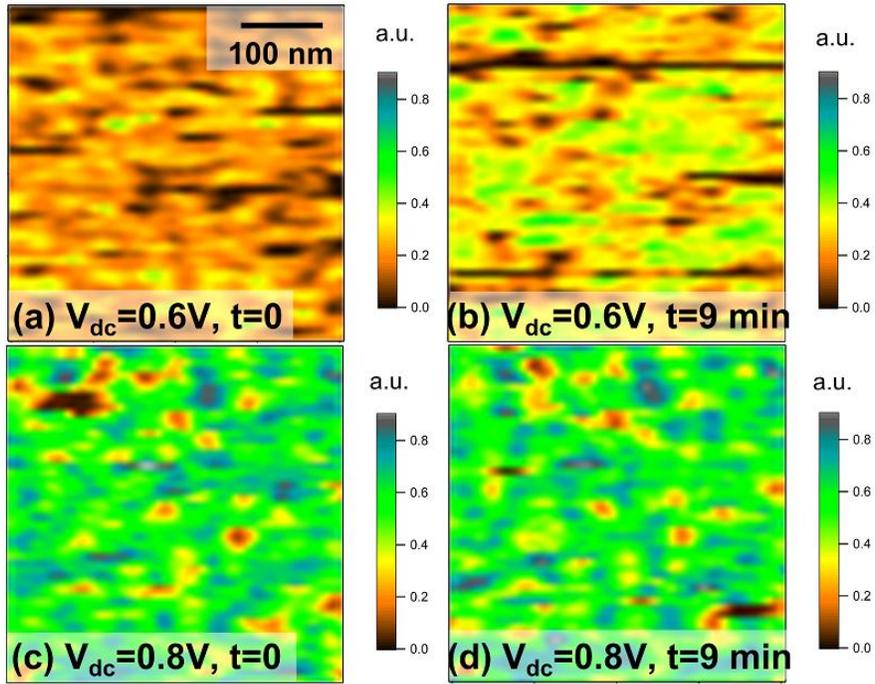

**FIG.S5**

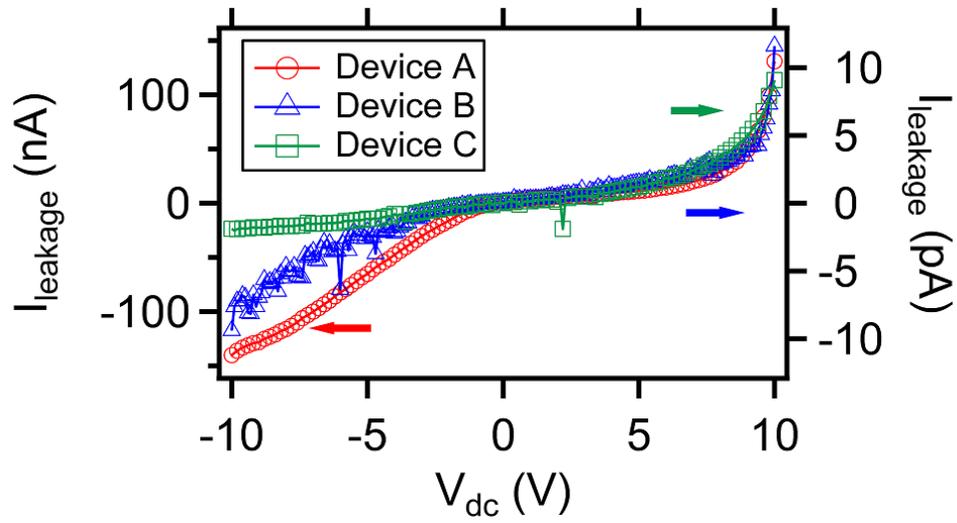

**FIG.S6**

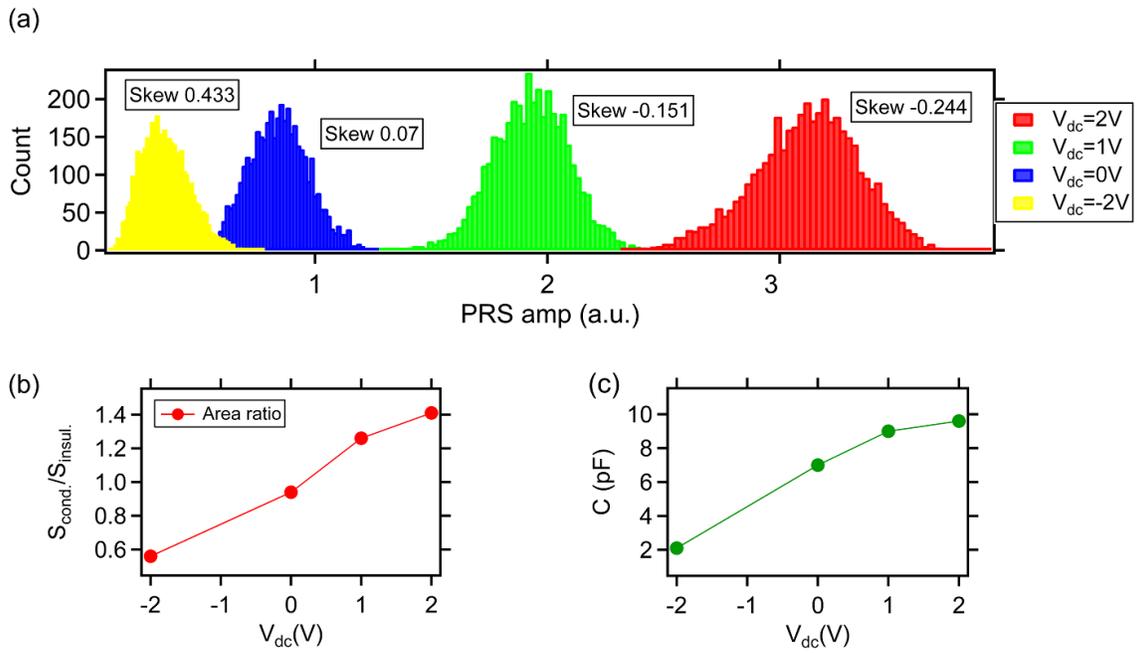

**FIG.S7**